# R&D around a photoneutralizer-based NBI system (Siphore) in view of a DEMO Tokamak steady state fusion reactor

A. Simonin[a], Jocelyn Achard[i], K. Achkasov[a, j], S. Bechu[h], C. Baudouin[a], O. Baulaigue[a], C. Blondel[e], JP. Boeuf[f], D. Bresteau[e], G. Cartry[j], W. Chaibi[b], C. Drag[e], H.P.L. de Esch[a], D. Fiorucci[a,b], G. Fubiani[f], I. Furno[c], R. Futtersack[f], P. Garibaldi[a], A. Gicquel[i], C. Grand[a], Ph. Guittienne[l], G. Hagelaar[f], A. Howling[c], R. Jacquier[c], M.J. Kirkpatrick[k], D. Lemoine[d], B. Lepetit[d], T. Minea[g], E. Odic[k], A. Revel[g], B.A. Soliman[m], P. Teste[k].

[a] CEA, IRFM, F-13108 St Paul lez Durance, France.
[b] Laboratoire ARTEMIS-UMR CNRS 7250, Cote d'Azur Observatory, Nice, France.
[c] Ecole Polytechnique Fédérale de Lausanne, Centre de Recherches en Physique des Plasmas CH-1015 Lausanne, Switzerland
[d] Université de Toulouse, Université Paul Sabatier, Laboratoire Collisions Agrégats Réactivité, IRSAMC(LCAR); CNRS, UMR 5589, F-31062 Toulouse, France
[e] Laboratoire Aimé-Cotton (LAC), CNRS, université Paris-Sud, école normale supérieure de Cachan, bât. 505, F-91405 Orsay cedex.
[f] Laboratoire Plasma et Conversion d'Energie, LAPLACE, P. Sabatier University, F-   Toulouse, France.
[g] Laboratoire de Physique des Gaz et des Plasmas, LPGP,UMR 8578 : CNRS- Paris-Sud University, F-91405 Orsay, France.
[h] LPSC, Université Grenoble-Alpes, CNRS/IN2P3, F-38026 Grenoble France
[i] Laboratoire des Sciences des Procédés et des Matériaux; LSPM, CNRS-UPR 3407 Université Paris 13, F-93430 Villetaneuse ; Labex, SEAM, PRES Sorbonne Paris Cité, France
[j] Laboratoires de Physique des Interactions Ioniques et Moléculaires ; PIIM ; Aix-Marseille University, CNRS, UMR 7345, 13013 Marseille, France
[k] Laboratoire Génie Electrique, Electronique de Paris, GeePs, CNRS UMR 8507, CentraleSupélec, UPSud Université Paris-Saclay and UPMC Sorbonne Université, 91192 Gif-Sur-Yvette, France
[l] Helyssen Sàrl, Route de la Louche 31, CH-1092 Belmont, Switzerland.
[m] Accelerators and Ion sources Department, Nuclear Research Center (NRC), Atomic energy Authority. P.O. Box: 13759 Inchas, Atomic Energy, Cairo, Egypt.

**Abstract**. Since the signature of the ITER treaty in 2006, a new research programme targeting the emergence of a new generation of Neutral Beam (NB) system for the future fusion reactor (DEMO Tokamak) has been underway between several laboratories in Europe. The specifications required to operate a NB system on DEMO are very demanding: the system has to provide plasma heating, current drive and plasma control at a very high level of power (up to 150 MW) and energy (1 or 2 MeV), including high performances in term of wall-plug efficiency ($\eta > 60\%$), high availability and reliability. To this aim, a novel NB concept based on the photodetachment of the energetic negative ion beam is under study. The keystone of this new concept is the achievement of a photoneutralizer where a high power photon flux (~3 MW) generated within a Fabry Perot cavity will overlap, cross and partially photodetach the intense negative ion beam accelerated at high energy (1 or 2 MeV). The aspect ratio of the beam-line (source, accelerator, etc.) is specifically designed to maximize the overlap of the photon beam with the ion beam. It is shown that such a photoneutralized based NB system would have the capability to provide several tens of MW of $D^0$ per beam line with a wall-plug efficiency higher than 60%. A feasibility study of the concept has been launched between different laboratories to address the different physics aspects, i.e., negative ion source, plasma modelling, ion accelerator simulation, photoneutralization and high voltage holding under vacuum. The paper describes the present status of the project and the main achievements of the developments in laboratories.

**Keywords:** Ion source, plasma modelling, Photodetachment, Fabry-Perot cavity, Electrostatic accelerators, negative ions, neutral beams, high voltage holding in vacuum, DEMO.

# INTRODUCTION

The construction of ITER raises the question of the next step toward a real fusion power plant, the DEMO Tokamak, which should be the first fusion reactor to demonstrate electricity production over long periods. In order to produce the required 500 MW of electricity coupled to the grid [1,2], the fusion reactions have to provide 1.5 GW of thermal power. Additional heating systems must be implemented in the reactor system in order to provide the initial plasma heating to enter the burn phase; the heating power required for a pulsed machine with no (or low) current drive is 50 MW whereas for a quasi-steady state reactor (pulse length of 300h) with a high current drive level [3,4,5] no less than 150 MW is required. Two heating systems are so far being considered for DEMO [1]: the neutral beam injection (NBI) system which injects high-energy neutral atoms into the plasma core, and the electron cyclotron resonance heating (ECRH) system which accelerates electrons in the plasma core via electromagnetic waves (f~170 GHz). The overall power efficiency of these plasma heating systems becomes an important parameter for a power plant as it directly impacts on the net electrical power produced by the reactor and the electricity cost. While with the present device (the ITER NBI system), the efficiency is lower than 30% [6], a tolerable electricity cost produced by a fusion reactor (DEMO) requires a global wall-plug efficiency higher than 60% [7].

It is clear that the NBI system remains an essential part of a reactor device, with very stringent requirements, such as the realisation of powerful high-energy beams, i.e., several tens of MW of neutral power at an energy ranging between 1 (for the pulsed machine) and 1 to2 MeV (for the long pulse reactor) [1], high wall-plug efficiency (> 60%) and a small footprint on the reactor environment. The achievement of such a performance requires considerable R&D effort in parallel to the ITER construction.

Presently, the NBI systems' negative ion beam neutralization is achieved by the stripping of the extra electron from negative ions through collisions with gas injected into the neutralizer cell. It is a simple and reliable method, but the neutralization efficiency is modest (around 55%) and the amount of gas injected both in the source and neutralizer leads to a high background gas density within the accelerating channel such that in the ITER-NBI system, about 30% of the negative ions being accelerated are lost [8] due to molecular collisions, thus contributing to the poor injector efficiency.

The neutralization of high-energy negative ions by photodetachment (photoneutralization) is an attractive alternative to beam neutralization by a gas target for several reasons [9,10]: a potentially high beam neutralization rate (higher than 80%), a complete suppression of the gas injection in the neutraliser, which amounts to 80% of the total gas injected along the ITER-beamline [6], reduction of stripping losses in the accelerator and reduction in parasitic particles (electrons, neutrals, positive ions) generated inside and outside the accelerator which load the beamline components.

On the other hand, it is a challenging method which requires significant R&D efforts before considering its implementation on an NBI system; indeed, due to the low photodetachment cross-section (about 4 $10^{-21}$ m$^2$) [9] and a short interaction time, a high photon flux (higher than 10MW of laser power) is required to attain a high neutralization rate (>80%) [10,11].

Suitable light flux amplification can be achieved in an optical Fabry-Perot cavity set across the ion beam, to a factor high enough to achieve around 3 MW of photon power, within realistic experimental parameters [10]. This would require a cavity finesse of 10000 fed by an amplified single-mode CW laser in the 1 kW range. Since the diameter of the light beam, of several centimeters, would be much larger than the half-wavelength (half a micron), only a very slight tilting of the laser beam, with respect to orthogonality to the ion beam, will be enough to make the ion flight smooth out the node and antinode structure. The duplication of such cavities (3 to 4) along the D⁻ beam within the neutralizer cell would result in the required high photon flux (higher than 10MW).

This new injector concept called Siphore, is a break from the conventional devices based on gas neutralization; it requires dedicated R&D in different fields of physics in order to cover all the different injector aspects: ion source& accelerator (this paper), photodetachment physics [9], high power Fabry-Perot cavity [10,11], caesium-free negative ion formation [12], and R&D on high voltage holding under vacuum [13,14].

After a presentation of the injector concept (chapter 1), this paper highlights the accompanying R&D in the different laboratories in France: chapter 2 describes the developments around a dedicated negative ion source which fits with the Siphore; chapter 3 is devoted to 2D & 3D simulations of the ion beam in the accelerator, photoneutralizer and energy recovery system; chapter 4 presents the R&D and advances in photoneutralization, and chapter 5 describes an ongoing project targeting the development of a new high voltage bushing for Siphore.

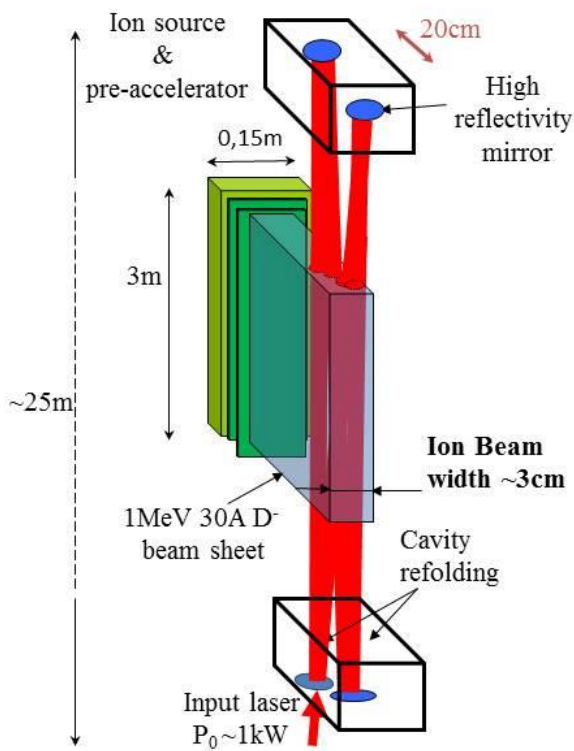 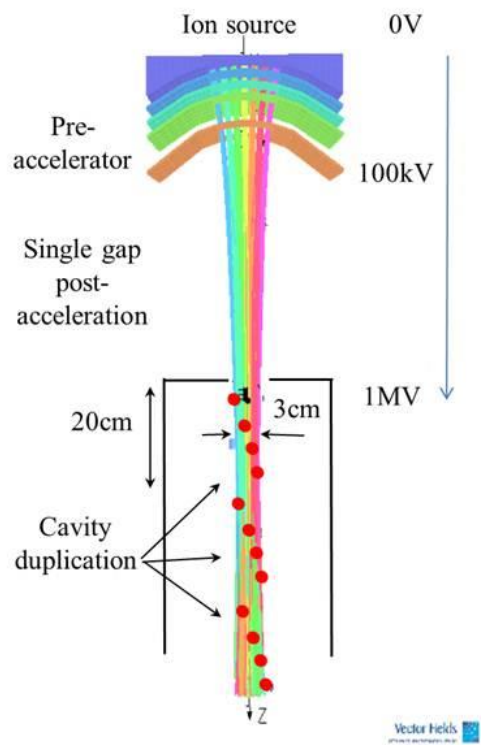

Figure 1.1: Topology of a photoneutralization based injector

Figure 1.2: 3D simulation of the D⁻ beam optics: pre-acceleration at 100 keV of seven beamlets in curved grids and post acceleration in a single gap at 1MeV.

# 1-SIPHORE CONCEPT

The interaction of the photon beam with the high energy ion beam has to be maximized by a specific design and aspect ratio of the beam line (see Fig. 1.1); the source & accelerator have to provide a thin and intense ion beam sheet (~30 A of D$^-$ at 1 or 2 MeV) in the neutralization region which has to be entirely overlapped and crossed from side to side by the photon beam [10,11]. As a direct consequence, this negative ion beam sheet is provided by an ion source and accelerator with dimensions: ~3 m high, ~15 cm wide. The concept significantly differs from conventional NBI systems. Figure 1.2 shows a 3D simulation (top view) of the accelerator, the negative ions are pre-accelerated up to 100keV, and merged into a single macro beam in the post-acceleration gap, in order to form a thin ribbon beam. The photoneutralizer is an equipotential cell held at 1MV, the stripped electrons at an energy of 270eV (1MeV/(m$_D$-/m$_e$-)) released by the photodetachment are trapped by the 1MV potential well; they are swept from the ion beam and dumped on the metal wall of the photoneutralizer cell. The secondary plasma density formed by the interaction of the charged particles (D$^-$ beam & electrons) with the background gas (~5mPa) is low (n$_e$~10$^{16}$ m$^{-3}$) leading to a positive ion (D$^+$) current leakage from the apertures in the mA range (I$_{D+}$<10mA).

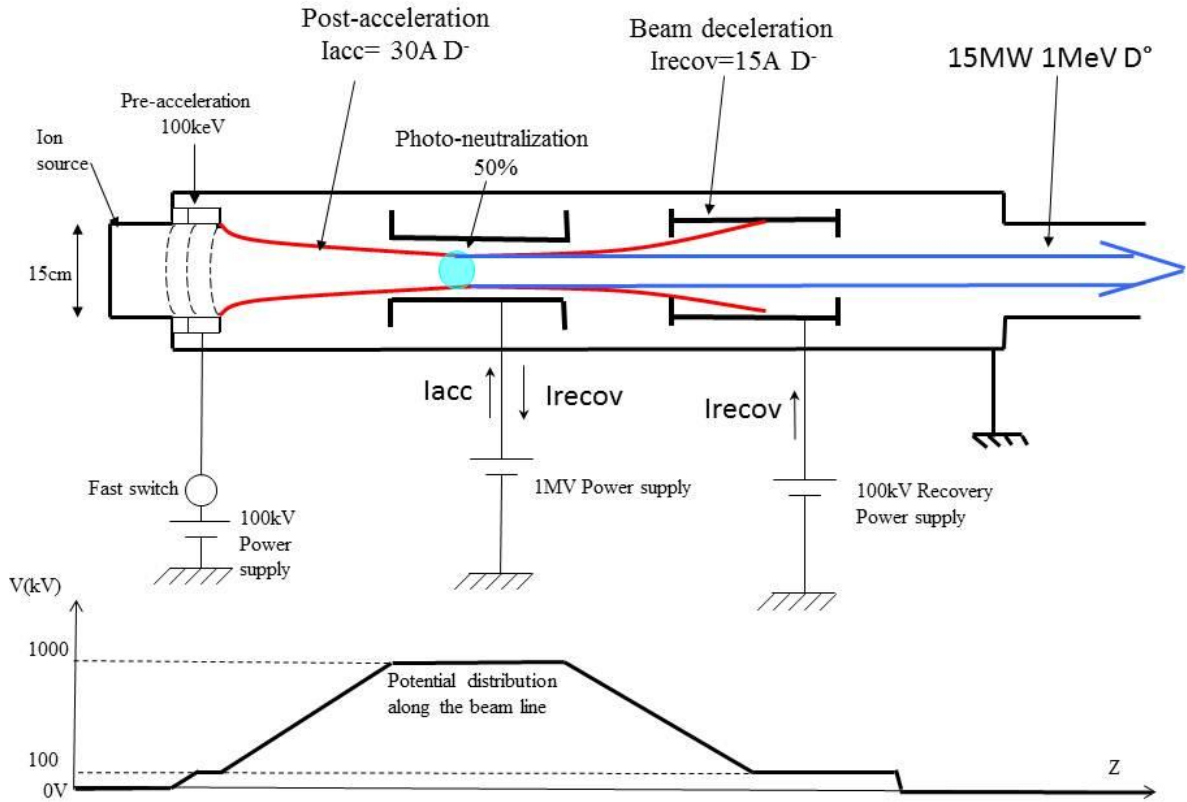

Figure 1.3: Principle of a 1MeV Siphore injector (Top view), and potential distribution along the beam line. The source is grounded and the photoneutralizer is held at +1MeV.

At the photoneutralizer exit, the 1 MeV non-neutralized fraction of negative ions (D⁻) are decelerated down to a low energy (<100 keV) and collected on the recovery electrode; the electrostatic force being conservative, the load (current) of the high voltage (1MV) power supply is then reduced. In this way high injector efficiency is reached even with incomplete photoneutralization. In summary, the injector concept is composed of a SIngle gap 1MeV accelerator, followed by a PHOtoneutralizer and a Recovery Energy system, leading to the acronym SIPHORE.

At this point we can point out the major difference between Siphore with the 1 MeV MAMuG conventional accelerator concept (ITER type) [6] which consists of five grids following the extractor, each at a potential 200 kV above that of the preceding grid, each having 1280 apertures aligned with the apertures in the other grids to form the 40 A D⁻ beam. With the gas neutralizer, ITER has chosen the MAMuG concept due to a lower electron leakage than the Single gap accelerator.

Figure 1.1 also shows that a complete illumination (overlap) of the D⁻ beam can be achieved by a cavity refolding (~3 to 4 refoldings) which steers the photon beam transversally. We note that it is a compact system in the direction parallel to the ion beam (see Fig. 1.1) with a width of 20cm. The photodetachment area is located at the ion beam waist (Fig. 1.2) close to the photoneutralizer entrance where the divergence is minimal ($\theta$<5mrad). Moreover, as the space charge force decreases with increasing beam energy (1 or 2MeV), several adjacent, identical and independent cavities along the beam waist region can be implemented increasing the effective laser power (and the overall neutralization rate), i.e., three adjacent cavities of 3MW each providing a neutralization rate of 50% leading to a total neutralization rate of 87.5% at 1MeV.

**1.1 1MeV injector efficiency evaluation**: For this assessment, we assume:

-i) A D⁻ beam ($I_{D^-}$ =30A) accelerated at 1MeV ($P_{D^-}$ =30MW) with N=87.5% of photodetachment provided by three adjacent Fabry-Perot cavities, and a recovery electrode which collects the non-neutralized fraction $I_{recov}=I_{D^-}*(1-N)$= 3.75A of D⁻ at $E_{recov}$=100 keV => $P_{recov}$=0.37 MW

-ii) T=80% transmission of the 1MeV neutral beam (including beam re-ionization in the duct)

-iii) Other electrical power losses: $P_s$=1MW to supply the source and pre-accelerator, $P_{sub}$=2MW for the subsystems (pumping and cooling systems), and $\eta_{ps}$=90% efficiency of the 1MV power supply

-iv) The CW single mode laser power feeding the optical cavity being in the range of a few kW is negligible.

The wall plug efficiency $\eta_{wp}$ is the ratio of the neutral power injected in the plasma by the total electrical power consumed to produce this neutral beam:

Eq.1: $$\eta_{wp} = \frac{P_{D^-} * N * T}{\frac{P_{D^-} * N}{\eta_{ps}} + P_{recov} + P_{sub} + P_s} = 0.64$$

The system would provide 21MW of neutral beam at 1MeV injected in the plasma core, for an overall injector efficiency close to 64%.

It is worth noting that the 60MW 1MV power supply which provides 17MW of 1MeV D° beam on ITER could supply two 30A "Siphore beamlines" implemented in parallel within the same vacuum tank to provide 42MW (2x 21MW) of D° at 1MeV.

## 1.2 Towards a 2 MeV high efficiency NBI system :

Neutral beams at 2MeV have significant advantages with respect to 1MeV beams; indeed, the increase by a factor two of the neutral power with the same extracted D$^-$ current (reduction of the number of beamlines), and a significant increase of the non-inductive current drive (around 20%) [15] by respect to a 1MeV beam.

The Siphore concept allows an increase in beam energy up to 2MeV via a tandem configuration (see Fig. 1.4) where the ion source and pre-accelerator are held at -1MV, separated from the neutralizer at +1MV by a central grounded electrode. The 1MV electrical set up (1MV power supplies, bushing, etc.) is symmetrical with opposite polarities on both sides of the grounded electrode. The source and pre-accelerator are suspended under vacuum and powered through the -1MV bushing; they provide the ion beam sheet which is then post-accelerated in two steps towards the photoneutralizer at +1MV. The unneutralized negative ions (at 2 MeV) will be deflected out from the neutral beam, decelerated up to 100keV and collected on a cooled target polarized at -900kV.

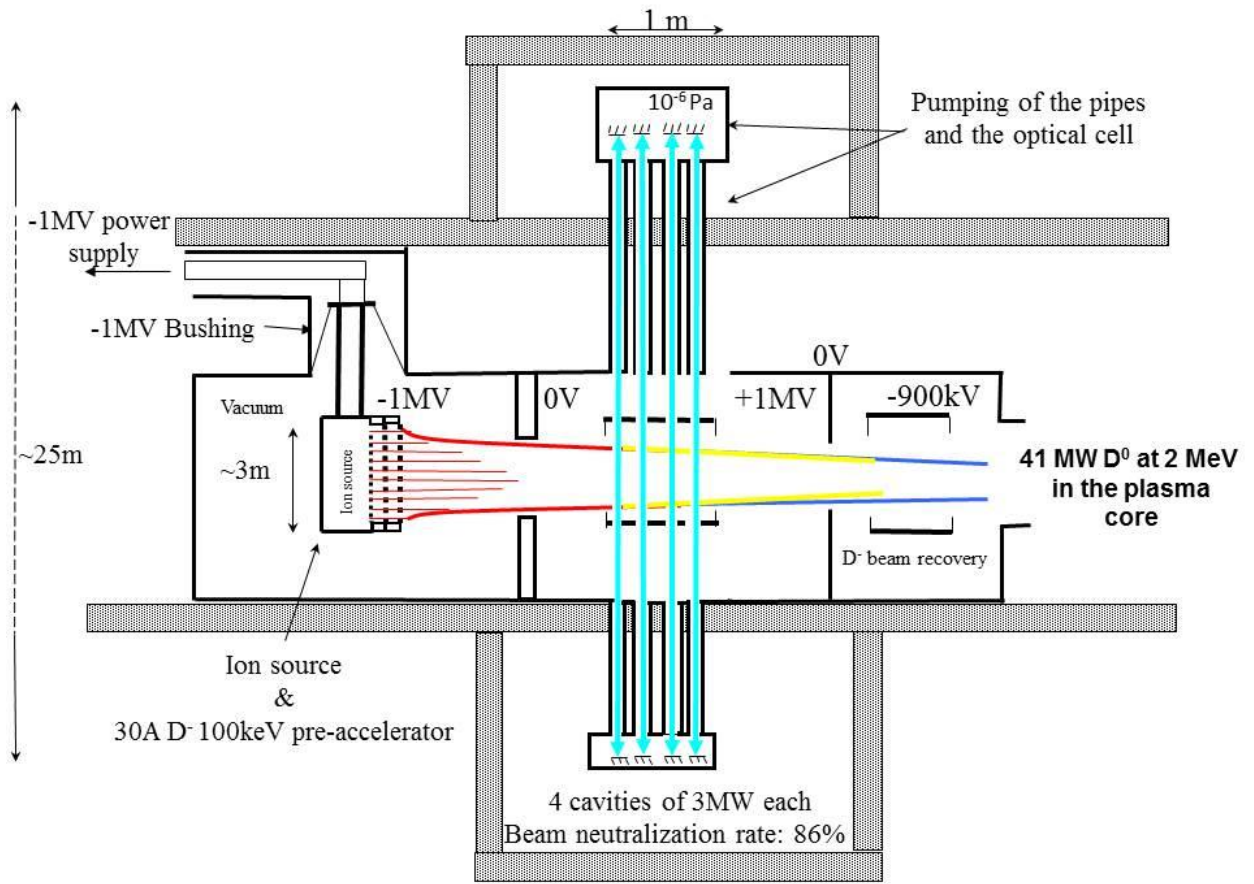

Figure 1.4: Side view of a 2MeV Siphore injector, with the source at -1MV and the photoneutralizer at +1MV; the ion source and pre-accelerator are sustained under vacuum polarized at -1MV and supplied by a vertical bushing. The photoneutralizer and recovery cells are sustained under vacuum by lateral Bushings (not shown on the figure) at +1MV and -900kV respectively.

The photodetached fraction being a function of the time of flight of the negative ion in the photon beam, it decreases by a factor (1-exp(-Ln(2)/√2)) at 2MeV leading to only 39% of neutralization ratio (instead of 50% at 1MeV) with a 3MW photon beam. The duplication of several cavities along the ion beam will compensate for this decrease (4 cavities ⇔ 86% of theoretical neutralization rate). The system would provide 41MW of neutral beam at 2MeV injected in the plasma core, for an overall injector efficiency close to 67%.

At such a high power density, further studies have to be addressed to evaluate the heat load within the duct and on the first wall components facing the beam due to the plasma shine through. It is important to notice that 120 MW of heating power concentrated in only three neutral beams could, in case of beam fault, have a significant impact on the reactor availability.

We note that the power losses due to beam re-ionization in the duct of the 2MeV neutral beam is the same range that at 1MeV due to a decrease by a factor two of the cross section: indeed, the cross section for the formation of positive ions in collision of the energetic D°(f) with gas ($D_2$), D°(f) + $D_2$ => $D^+$(f) + $D_2$ + $e^-$, is maximum at 20keV and monotony decreases at higher energy. At 1MeV, the cross section is $\Gamma(1MeV)= 3.68 \cdot 10^{-21} m^2$, and at 2MeV, $\Gamma(2MeV)=1.91 \cdot 10^{-21} m^2$. Moreover, to avoid beam interception (heavy heat loads) within the duct, the beam divergence and aberrations have to be minimized.

We can see on figure 1.4 that the optical components (cavity mirrors) are located far away from the injector (~15m away) and outside the nuclear island of the beam line and Tokamak in order to prevent them as much as possible from a rapid degradation by pollutants (plasma-gas-metallic sputtering, radiation (neutrons)) and to facilitate an easy maintenance. Intermediate active pumping cells and neutron absorbing materials would contribute to keep the optical cells under high and clean vacuum conditions ($P_{optical\ cell}$~$10^{-6}$Pa).

## 2- ION SOURCE DEVELOPMENT FOR SIPHORE

Conventional negative ion sources [16] used for NBI systems are based on either filaments or on radio frequency (RF) heating. All make use of caesium to produce sufficient amounts of negative ions. Conventional RF sources are ICP (Inductively Coupled Plasma) driven ion sources where several RF plasma generators (called Drivers) at the back of the source produce a hydrogen or deuterium plasma which bombards through a transverse magnetic field (called filter field) the first accelerator electrode, the Plasma Grid (PG), where negative ions are formed and extracted. The filter field acts as a magnetic barrier which cools down the hot electrons generated within the drivers and prevents a high destruction rate of the negative ions formed on the PG. The source has to produce high and uniform current density ($J_{D^-}$ = 250 A.m$^{-2}$ of extracted current density with ±10% of homogeneity) over a large area (plasma grid surface of ~ 0.8 m$^2$ on ITER) at a low operating pressure (p < 0.3 Pa) to limit the negative ion losses by stripping reactions in the accelerator. Past experiments of source prototypes [16,17] and plasma modelling [18-22] have shown that an EXB plasma drift occurs along the vertical axis and leads to significant plasma inhomogeneity (along the vertical axis) in the extraction region. As a direct consequence, this source concept with the transverse filter field which leads to a significant plasma inhomogeneity along the vertical axis is not suitable for a long (~2.5m height) and thin (~15cm wide) Siphore ion source.

The Cybele ion source [23] is a tall and narrow ion source (see Fig. 2.1-a) with a rectangular aspect ratio that is particularly relevant to Siphore; its plasma source dimensions are: height 1.2m, width 15cm, depth 20cm. Cybele is a filamented plasma source in which 5 sets of 3 tungsten filaments are used as cathodes (Vcathode=-70V) along the source vertical axis. The filaments supply the plasma core with primary electrons along the vertical axis (see Fig. 2.1-c); the filament current ranges from 200 to 1500A (power up to 100kW); the source walls are connected to the ground potential. A uniform magnetic field parallel to the source vertical axis is generated by two lateral coils sitting on opposite sides of an iron rectangular frame which surrounds the source (see Fig. 2.1-b). The two coils generate magnetic fields in the opposite direction inside the iron structure (see Fig. 2.1-b). It is the leakage field between the two coils that then fills uniformly the plasma source volume. The magnetic field intensity within the whole plasma volume can be tuned between 0 and 7 mT by adjusting the DC electric current in the coils.

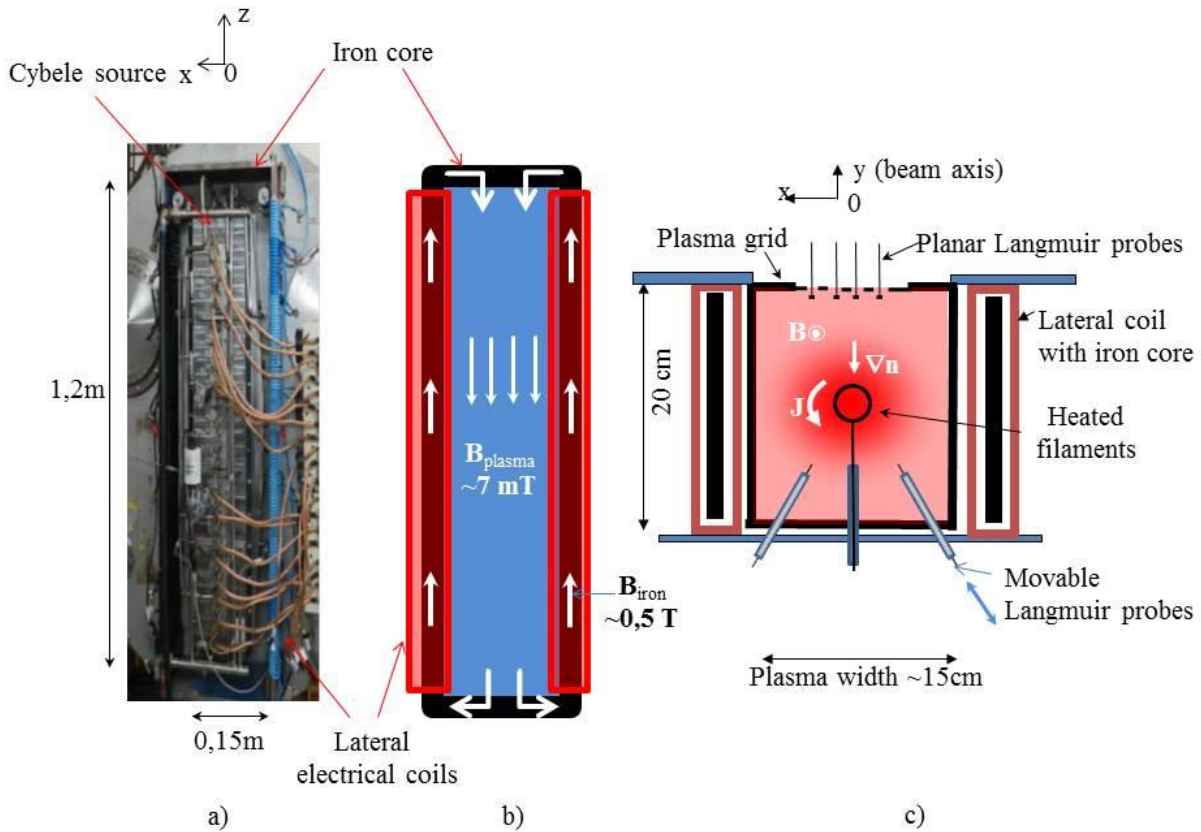

Figure 2.1: a) Photo of the Cybele source (back face) with the surrounding iron core and lateral coils; b) Schematic of the magnetic set up surrounding Cybele; c) Horizontal cross section of the source ; J is the plasma flow rotating around the cathode, and $\nabla n$ is the gradient of the plasma kinetic pressure.

## 2.1 Experimental results from the Cybele source [24]:

Figure 2.2 left shows radial profiles of plasma density ($n_e$) and effective electron temperature ($T_e$) for a power input of 30 kW with different magnetic field intensities. The Langmuir probe (see Fig. 2.1-c) scans the plasma 9 cm from the source wall to 2 cm inside the source (the plasma centre) where primary electrons of 70 eV are emitted. We observe that the plasma density increases in the source centre with the field intensity at up to 5.3 mT; this effect is due to the increase of the electron confinement by the magnetic field (the Larmor radius of the 70eV electron is only a few mm), such that the plasma density increases from $0.7 \times 10^{17}$ m$^{-3}$ up to $4.8 \times 10^{17}$ m$^{-3}$. Figure 2.3 shows the electron temperature, we note a better plasma cooling with increasing field intensity.

The floating potential (Vf) is -48 V in the plasma centre close to the filament suggesting the presence of hot electrons. Close to the source wall Vf is weaker (-14 V for B=3 mT).

We note in figure 2.4 a negative plasma potential for the two magnetic field intensities of 3 and 6.5 mT which results from the fact the electron radial mobility is lower than that of the ions. With the decrease of the radial electron current diffusing across the B-field line scaling as ~1/B (Bohm diffusion [25]), the plasma potential becomes more negative to maintain the electron transport towards the source wall. At low magnetic fields, the plasma potential is nearly uniform (-10 V at 3 mT) in the plasma bulk, while at higher field intensity (6.5 mT), a radial electric field appears likely due to a diamagnetic effect; this radial electric field could lead to rotating instabilities with an E x B drift [26].

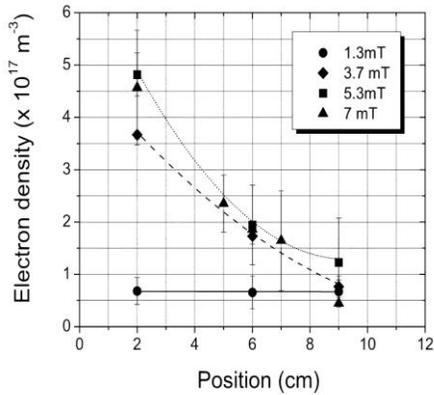
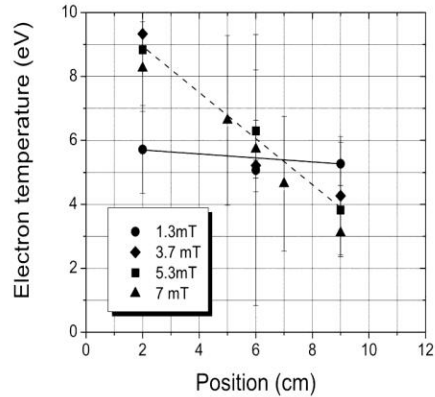

Figure 2.2: Radial profiles of electron density (ne) and effective electron temperature (Te) at different magnetic field intensities (1.3, 3.7, 5.3 and 7 mT). The source center is at 10 cm.

Figure 2.3 presents the variations of floating and plasma potentials from the source wall up to the centre at 6.5 mT and 3 mT (P$_{filaments}$ = 30 kW and source pressure 0.26 Pa).

The vertical and transverse plasma density distributions measured with the planar probes on the plasma grid are shown in figures 2.5 and 2.6 for B = 7mT, P$_{filaments}$ = 30kW and P=0.26 Pa. Along the source vertical axis (Fig. 2.5) the plasma density is nearly uniform but rapidly decreases over the last 10 cm at the source extremities due to the electron leak in the direction parallel to the B-field lines (due to the high electron mobility in the direction parallel to B).

Figure 2.6 highlights an asymmetry in the transverse (right-left) plasma parameters: on the left side, the plasma has a higher density (by a factor of 2), higher electron temperature and floating potential than on the opposite side (right). An inversion of the magnetic field in the source induces the same asymmetry on the opposite direction.

A scan of the source filling pressure shows that at a pressure lower than 0.25 Pa, the plasma density decreases and a saturation is observed above. A scan of the input power up to 100 kW at 0.2 Pa (with B = 7 mT) shows a linear increase in plasma density from $2 \cdot 10^{17} m^{-3}$ up to $8 \cdot 10^{17} m^{-3}$; above this, the plasma density saturates.

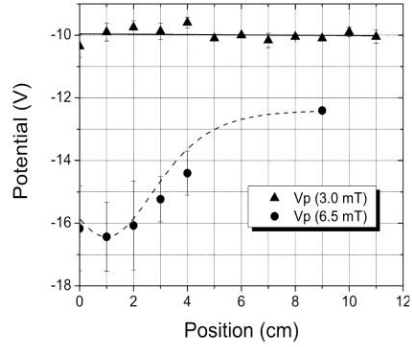

Figure 2.4: Radial plasma potential distribution for two magnetic fields (3 and 6.5mT) from the source centre (position 1cm) up to the edge (12cm).

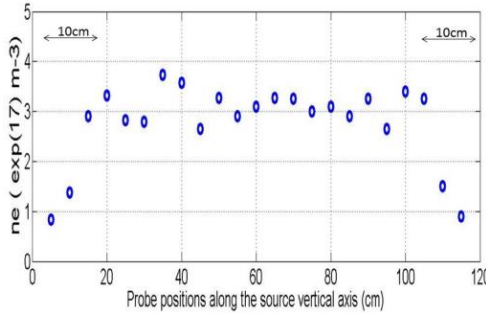

Figure 2.5: Plasma density along the vertical axis

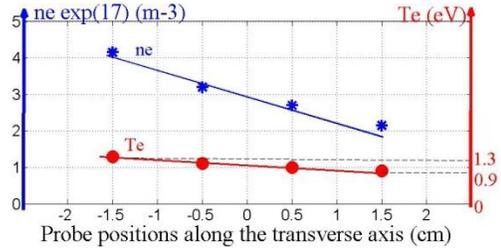

Figure 2.6: Transverse plasma profiles

## 2.3) 2D modeling of the Cybele source

Simulation of the Cybele source operation with filaments has been performed with the PIC MCC (Particle-In-Cell Monte Carlo Collisions) model described in Refs. [26-31]. The PIC MCC model is two-dimensional in the plane perpendicular to the magnetic field. As in Refs.[26-31], a discharge in pure $H_2$ is considered and no plasma chemistry is included in the model. A complete set of electron-$H_2$ collision cross-sections, charge exchange collisions of positive ions, and electron-ion Coulomb collisions are taken into account. Since it is practically impossible, with PIC simulations, to simulate steady state situations with plasma densities as large as $10^{18}$ $m^{-3}$, which are expected in Cybele, we consider smaller plasma densities (on the order of $10^{15}$ $m^{-3}$) and use a scaling factor. The Debye length (and sheath lengths) in the simulation are therefore much larger than in the real plasma and care must be taken in the interpretation of the simulation results (especially in the presence of non-classical collisional plasma transport or plasma turbulence). The simulations have been performed in a cylindrical geometry, with a chamber diameter of 14 cm and a column length L of 1 m.

Figures 2.7 shows the radial distribution of the normalized, time averaged plasma density for two values of the magnetic field, 3 mT and 5 mT calculated for a hydrogen pressure of 0.27 Pa. The densities are represented by a dotted line around the filament (position 0) because of the approximate representation of the filament and sheath around the filament in the model. The unit for the plasma density in Fig. 2.7 is $2\times10^{18}$ m$^{-3}$ for 5 mT and slightly below for 3 mT. These values are obtained for a total (scaled) current of around 500 A, corresponding to a power of 35 kW (500 A x 70 V).

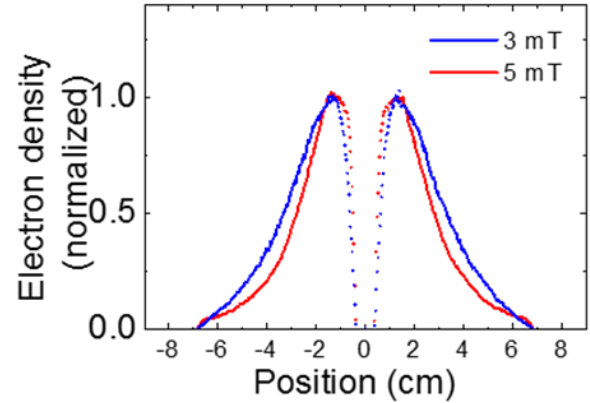

Figure 2.7: Radial distribution of the time averaged, normalized plasma density for two values of the magnetic field (3 and 5 mT) and a pressure of 0.26Pa.

The plasma density in the simulation exhibits some instabilities as can be seen in Fig. 2.8 with a normalized plasma density, which shows the 2D distribution of the plasma density at a given time of the simulation, for a magnetic field of 5 mT. The 2D distribution of the time averaged plasma density is shown for comparison in Fig. 2.9. The instabilities seen in Fig. 2.8 are rotating over a time on the order of 5 μs. This corresponds to a rotation velocity around $4\times10^6$ cm/s at a radius of 3 cm, i.e. close to the critical ionization velocity of hydrogen [28].

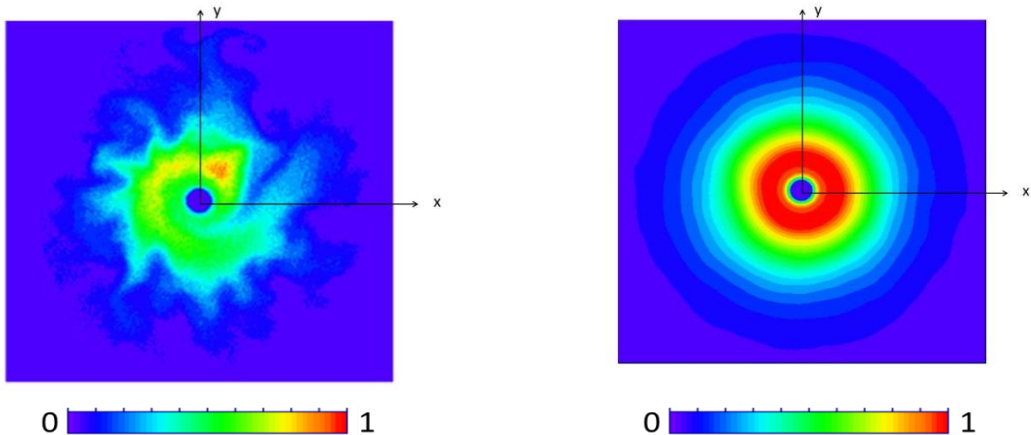

Figure 2.8: Distribution of the plasma density (normalized plasma density) at a given time for B=5 mT, p=0.26Pa (linear scale, unit $2\times10^{18}$ m$^{-3}$).

Figure 2.9: Time averaged distribution of the plasma density in the conditions of Fig. 2.8 (linear scale, unit $2\times10^{18}$ m$^{-3}$).

The increase of the plasma density in the source center that can be seen in the figures 2.7 and 2.9 is consistent with the experimental measurements (see Fig. 2.2) and is due to the magnetic confinement of the primary electrons (the Larmor radius of 70 eV electrons in a 3 mT field is about 5 mm); indeed, most of the ionization takes place in a region of a few cm radius around the filament (cathode). The time averaged electron temperature for the magnetic fields is shown in Fig. 2.10. This temperature is on the order of 10 eV around the filament and drops to about 1 eV next to the wall.

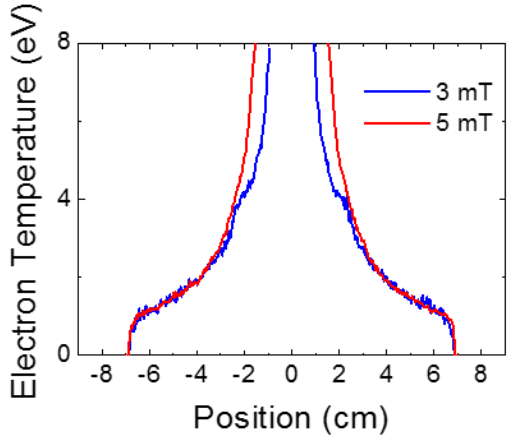 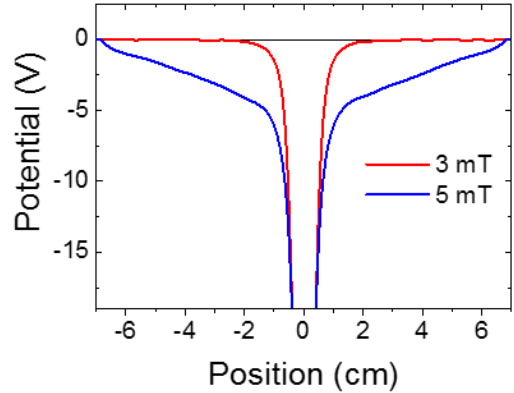

Figure 2.10: Radial distribution of the time averaged electron temperature.

Figure 2.11: Radial distribution of the time averaged plasma potential.

The radial profile of the time averaged plasma potential in Fig. 2.11 shows that at a low magnetic field (3mT), the plasma bulk is equipotential (E=0V/cm), while at a higher B-field (5mT), a time averaged radial electric field of about 1 V/cm appears in the plasma. This is consistent with the Langmuir probe measurements in Cybele (see Fig. 2.4).

Further simulations including negative ions ($n_{H^-} \approx n_{e^-}$) generated on the source wall have been performed. The principal properties of the plasma are not strongly modified by the presence of negative ions. The rotating instabilities may however be less important in the presence of non-negligible densities of negative ions because of the resulting increase in conductivity (ions are much less magnetized than electrons). Work is in progress to quantify this effect and its consequences on the plasma properties.

## 2.2 Negative ion production in Cybele

Experiments with filamented arc discharges have shown that the primary electrons emitted in the plasma centre along the source axis undergo a radial cooling (from 9eV in the source centre up to 1-2eV on the lateral walls) which should allow the production of negative ions in the extraction region (in the nearby of the accelerator). But, the tungsten evaporated by the heated filaments would contaminate the low work function metal surfaces where negative ions are formed.

Negative ion production in Cybele will take place with RF plasma generators (ICP or helicon) implemented at the source extremities (top, bottom) where the plasma leaks (see Fig. 2.6) in order to inject the plasma particles in the direction parallel to the magnetic field. The RF generators will be immersed in the source magnetic field generated by the surrounding electro-magnet. Two different drivers will be developed and tested on Cybele:

*-i) a conventional (ITER type) ICP driver* (1MHz , 10cm diameter) with a solenoid antenna is being tested on Cybele at a RF input power ranging from 10 to 50kW.

*-ii) Helicon plasma generator:* The ability to obtain high plasma density with high ionisation rate and much higher power efficiency than ICP generators makes Helicon sources an interesting candidate to feed Cybele with a dense and hot plasma along its central axis (like the filamented cathodes). For this purpose, a 10kW, 13.56MHz helicon plasma generator is under development at the CRPP-EPFL Lausanne. Although a single 10kW helicon generator will probably not achieve the relevant plasma density required for NB sources for future fusion grade reactors, the 10kW helicon source is an intermediate step towards larger powers, which will allow investigating the main technology and physics issues related to high power helicons.

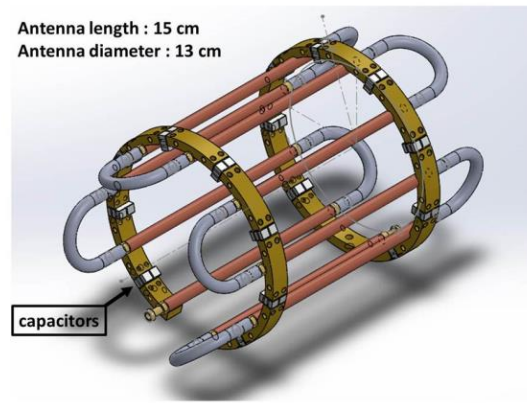

Figure 2.12: 9 leg cylindrical resonant network used for helicon excitation. The capacitors value is 3840pF to bring the m=1 resonance close to 13.56 MHz (with the screens).

In the past years at CRPP, a new type of helicon source was developed, which is based on the concept of a resonant birdcage network antenna [32-34]. A birdcage resonant antenna is shown in Fig. 2.12 using a CAD drawing of the antenna specifically designed to fit on Cybele. The antenna consists of conducting parallel legs distributed around a dielectric tube in a cylindrical configuration. Each leg is connected at both ends to its closest neighbours by capacitors. This structure can be seen in a first approximation as a parallel arrangement of L, C lumped elements, and presents a set of resonant frequencies corresponding to the normal modes of the structure. When excited at one of its resonant frequencies a strong, azimuthally sinusoidal, distribution of current amplitude is generated in the antenna legs, all these currents being temporally in phase. The RF fields generated by these current distributions fit well the helicon wave field structure, thus resulting in an efficient excitation of helicon waves. In a first phase before the test on Cybele, the helicon source will be installed on a vacuum chamber at CRPP. First studies will focus on important issues in the application as NBI plasma sources (hydrogen gas, low magnetic field operation…). Particular care will be devoted to study the important thermo-mechanical response of the high RF power plasma source.

*-iii) Development of Cs-free solutions for negative-ion production*:

The only up-to-date available scientific solution to reach the high $D^-$ negative-ion current required for fusion is the use of caesium. Studies conducted at IPP Garching on the ITER negative-ion source show that ITER requirements in terms of current density can be reached using this solution [35,36]. Caesium is injected in the negative-ion source and deposits on all surfaces in contact with the plasma. Deposition of caesium lowers the material work function and allows for high surface ionization efficiency upon impact of deuterium positive ions or atoms, and thus allows reaching high negative ion yields. The caesium method can be used in the Cybele negative-ion source to produce negative-ions. However, drawbacks to the use of caesium have been identified for the future fusion reactors. First, to obtain stable negative-ion currents over long shots, a continuous injection of caesium is required, leading to a high caesium consumption (~3 µg/s [37,38]). Second, caesium diffusion and pollution of the accelerator stage might cause parasitic beams and/or voltage breakdowns and imply a regular and restrictive maintenance in a nuclear environment. Therefore, a caesium-free solution would be highly valuable for future NBI devices. PIIM and LSPM laboratories are working on caesium-free solutions to produce negative-ions. Diamond which presents a negative electron affinity (conduction band above the vacuum level) is an attractive material.

In order to study negative-ion surface production in caesium-free plasmas in parallel to the development of Cybele, and in particular to study various layers of diamond materials, a dedicated experiment has been designed at PIIM laboratory [12, 39-44]. The aim is to evaluate and optimize alternative solutions to caesium prior to testing them in the Siphore negative-ion source.

The figure 2.13 presents yield measurements. Several carbon materials (in particular, several diamond layers) have been compared. Micro-Crystalline Boron Doped Diamond (MCBDD) and Micro-Crystalline non doped Diamond (MCD) as well as two different layers of Nano-Crystalline non doped Diamond (NCD) have been tested, all of them having been deposited at LSPM laboratory. Apart from diamond, HOPG and CFC (Carbon Fibre Composites used in the past as carbon tiles for tokamaks, here the C/C composite Sepcarb N11®) have been employed. The yields have been measured at different surface temperatures with the sample facing a mass spectrometer. The surface is negatively biased and the ions are self-extracted towards the mass spectrometer. It has been checked that this measurement with the sample surface normal to the mass spectrometer axis was representative of the global yield (at any angle). The striking point on this graph is the different behaviour between the diamond group of materials and the other carbon layers. While on carbon layers the yield mostly decreases with temperature, the yield on diamonds at first increases with temperature, reaches a maximum around 400-500°C and then decreases. There is on the average (depending on experimental conditions) an increase of the yield by a factor of 3-5 compared to room temperature.

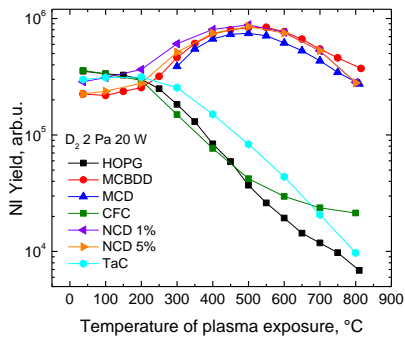

Figure 2.13: Negative ion yield versus surface temperature in low-pressure (2 Pa) deuterium plasma (RF power 20W). MCBDD, MCD, and NCD, stand for Micro-Crystalline Boron Doped Diamond, Micro-Crystalline Diamond, and Nano-Crystalline Diamond respectively. HOPG, CFC and TaC stand for Highly Oriented Pyrolitic Graphite (HOPG), Carbon Fiber Composites (CFC) and tetrahedral amorphous carbon (taC)

Raman measurements have been performed after plasma exposure for some of these materials [42-44]. The yield maximum observed for diamond materials has been attributed to electronic properties of the top surface that are favorable to surface ionization. These electronic properties are tuned by the sp3/sp2 hybridization phases ratio and by the hydrogen percentage [42-44]. The future work aims for better definition of the key surface state parameters leading to the high surface ionization efficiency. The diamond crystalline orientation will be studied using several kinds of diamond single crystals, the doping influence will be studied using boron and nitrogen as dopants.

## 3-ION BEAM ACCELERATOR

### 3.1- Siphore accelerator design and simulation

The negative ions are extracted from a thin and long ion source 3m high and 15 cm wide, which produces a uniform D$^-$ extracted density current of J$_{D-}$ = 250 A/m$^2$ (same range as the ITER ion source) over the extraction surface.

The accelerator will have to extract and shape the D$^-$ into a thin laminar beam (Fig. 1.2) which will be post-accelerated at high energy. For this purpose, the pre-accelerating grids have a transverse curvature radius (see Fig. 3.1 and 3.2) in order to merge and focus the beamlets in a single macro beam sheet in the post-acceleration stage (see Fig. 1.2).

Moreover, the stray electrons in the pre-accelerating stage have to be efficiently suppressed. For this purpose, they will be deflected out from the ion beam and dumped onto the grid metal surfaces by a vertical magnetic field diffusing from the ion source; a complete electron suppression at the pre-accelerator exit requires a field intensity ranging between 6 to 8 mT (see below); the grids do not contain any permanent magnets. In Fig. 3.1, we note that instead of cylindrical beamlets as used on the ITER NBI system, the grids are based on slot apertures, each slot is 0.9cm wide and 28cm high. Each grid contains seven adjacent slots leading to a high grid transparency close to 50%; the ITER grid transparency with cylindrical apertures and Sm-Co permanent magnets is 25%. The D$^-$ current extracted by one grid is 4.4 A. The entire accelerator extraction surface which faces the ion source will consist of several grids (3 grids per meter leading to about 13 A of D$^-$ per meter height) set up along the source vertical axis and spatially oriented in order to focus the neutral beam downstream within the duct of the Tokamak plasma chamber. The overall extracted D$^-$ current from a 3m height ion source ranges around 39A.

On Fig. 3.3 is depicted an estimate of the pressure profile within the pre-accelerator stage assuming 0.3 Pa of source pressure and a gas temperature of 300K. The stripping losses within the pre-accelerating channel is estimated to be 18%, leading to around 30-32 A of D$^-$ at the pre-accelerator exit (in the post-acceleration stage). Supposing a source pressure of 0.3Pa, the gas flow rate in the pre-accelerator is 1000 Pa l/s. With a pumping speed in the tank of S=2 10$^5$ l/s, the background pressure in the post-acceleration gap is 5 mPa leading to only 1.5 % of stripping losses for an acceleration to 1MeV. It is clear that photoneutralization leads to an important economy of the pumping system; the pumping speed of the ITER-NBI system based on gas neutralizer being 5 10$^6$ l/s [6] .

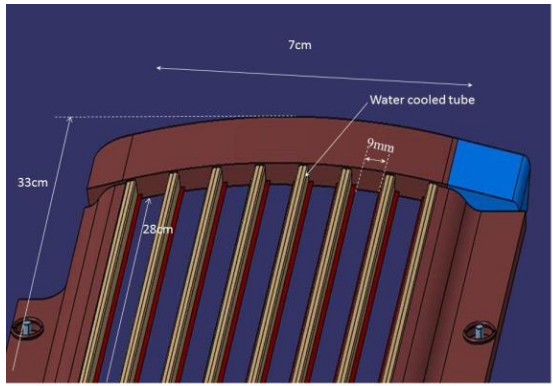

Figure 3.1: Engineering drawing (front view) of a pre-accelerating grid.

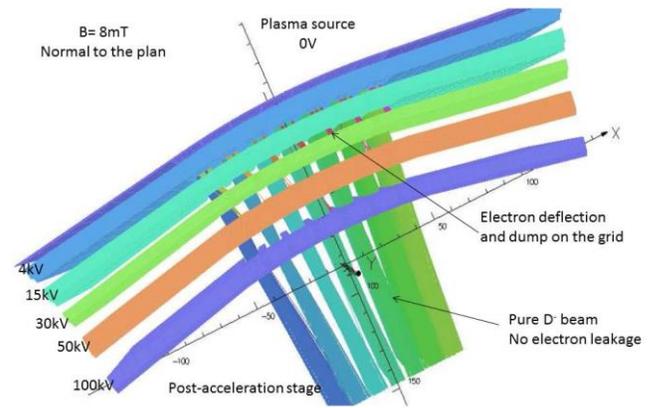

Figure 3.2: 3D simulation of the 100keV pre-accelerator

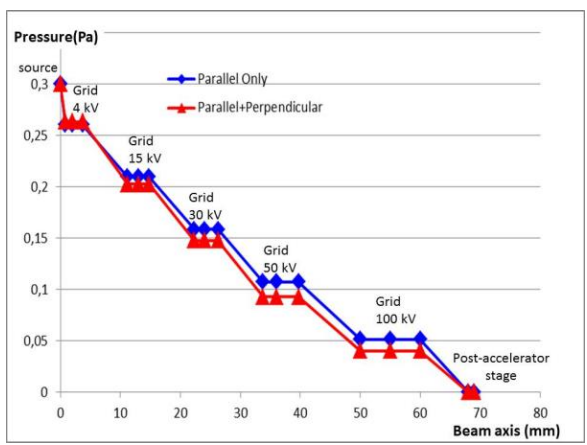

Figure 3.3: estimate of the pressure profile within the pre-accelerator stage; we note that the perpendicular conductance appears to be much smaller than the parallel conductance.

Fig. 3.2 shows a 3D simulation with the Opera-Scala code [45] of the electron suppression (deflection) in the 100 keV Siphore pre-accelerator by the magnetic field diffusing from the ion source. We note that the electron suppression is complete (no electron leakage in the post-acceleration stage), the major part of the co-extracted electrons are dumped on the first three grids (at 5, 15 and 30kV) for a magnetic field of 8mT. With the slot geometry, the thermal load is uniformly distributed along the tube height. Assuming one co-extracted electron per D$^-$ from the plasma source (i.e., ~4.4A of co-extracted electrons per grid), the thermal load on the downstream pre-accelerator grids is estimated by the electron current dumped by each grid (obtained from simulations):

$G_2$ at 5kV, $P_{G2}$ = 6kW;  $G_3$ at 15kV,  $P_{G3}$= 40 kW,  $G_4$ at 30kV, $P_{G4}$= 12 kW

Thermo-mechanical simulation of the grids has been done using the Ansys software. Simulations show that water cavitation occurs in the tubes when the thermal load is higher than 80 kW (per grid).

In parallel to the simulation with Opera, 3D simulation of the electron trajectories in this 100keV accelerating channel have been performed; the model [46] takes into account of the atomic processes occurring within the accelerating channel (interaction of particles with gas and metal surface), mainly the stripping reactions of negative ion collisions with the background gas and the secondary electron emission of the grid metal surfaces due to particle bombardment.

Fig. 3.4 shows the 3D trajectories of the co-extracted electrons with a low magnetic field (B=6 mT). We can note that the major part of the co-extracted electrons bombard the 30kV grid leading to a higher thermal load ($P_{G4}= V_{G4}*I_{G4}$=70 kW). As a consequence, there is a compromise to find between a high magnetic field (B ~ 8 to 10 mT) which efficiently suppresses the electrons on the low voltage grids (and reduces the thermal loads) but, which increases the plasma drift in the source (rotating instabilities).

Fig. 3.5 shows the 3D trajectories of the stripped electrons emitted within the accelerating channel; we note that they are intercepted by the last two grids (at 50 and 100 kV) and the suppression is nearly complete.

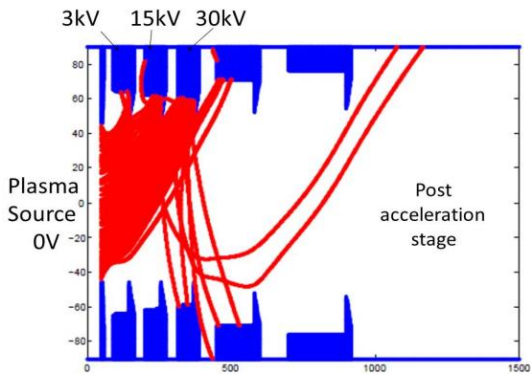
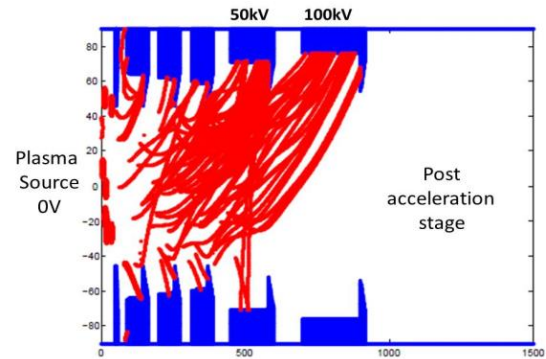

Figure 3.4: 3D trajectories of the co-electron in one pre-accelerating channel (top view of one slot) for a uniform magnetic field (normal to the plan of the figure) of 6 mT.

Figure 3.5: 3D trajectories of the stripped electron in one pre-accelerating channel for a magnetic field of 6 mT.

The uniform magnetic field in the pre-accelerator stage leads to an overall and uniform deflection the negative ions (ranging around 20mrad) at the pre-accelerator exit, which can be canceled by a lateral aperture displacement (δ~1mm) of the last grid (at 100 kV), or by tilting the set "ion source + pre-accelerator".

### 3.2 - Beam optics simulation in the photoneutralizer

The photoneutralization of the D⁻ high energy negative ion beam was primarily modelled, inside a 'vacuum' neutralizer system, following the Siphore concept described in the Introduction (Fig. 1.1). The meaning of the word 'vacuum' represents the very low level of the residual gas pressure filling the volume between neutralizer plates (less than 5mPa). Hence, the energetic beam particles interact exclusively with the laser photons trapped in each Fabry-Perot cavity, and the eventual development of a secondary plasma by the residual gas ionization is highly improbable. However, the 1 MeV $D^-$ beam stripping due to the interaction with the $D_2$ gas has been extensively studied for the ITER-like neutralizer [47].

Let us remember the main conclusions of the gas neutralization (i) the very efficient compensation of the $D^-$ beam space charge by the secondary plasma developed due to the ionization of the gas molecules by the energetic beam particles; (ii) the neutralization efficiency can reach the theoretical value, *i.e.*, ~56% conversion of $D^-$ in $D°$; (iii) the gas neutralization leads not only to one electron stripping but also to double stripping producing energetic (1MeV) $D^+$ at the neutralizer exit.

For the purpose of this study, ONAC (Orsay Negative ion ACcelerator) 3D code [48] has been adapted to describe Siphore photoneutralizer parameters such as plate potential, geometry and beam features given in Fig. 1.2. ONAC code developed at LPGP is based on the Particle-in-Cell (PIC) approach and it takes into account the space-charge of the beam solving 3D Poisson's equation.

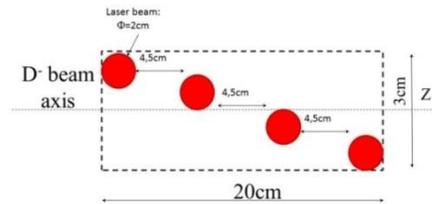

Figure 3.6: Top view of the simulation domain of one Fabry-Perot Cavity inspired by the photon trajectories depicted in Fig. 1.2. The cavity length is 20cm.

**Beam optic simulation with the duplication of two cavities:**

Beam simulations have been performed assuming a photoneutralizer with the duplication of two identical Fabry-Perot Cavities (FPC) implemented along the beam axis. The goal is to estimate the neutral beam emittance growth as function of the photoneutralization rate per cavity: one run was performed with 40% of photodetachment rate per cavity and a second run with 60%.

To evaluate the space charge effect on the emittance growth, a narrow pure parallel (zero divergence) 1MeV $D^-$ of 3 cm width 10cm height has been simulated which carries a current representing 30 A of $D^-$ for a linear beam of 3 m height (see the simulation domain on Fig. 3.6).

**Simulation with 40% photodetachment per cavity:** The simulation shows that the neutralization efficiency is very close to the expected theoretical value (~64%). As the photoneutralization process does not affect the particle momentum, than the $D°$ produced in the second FPC will continue with the small deflecting angle attained in the free space between the two laser cavities. This is visible on the emittance diagram shown in Fig. 3.7, which clearly presents two shapes. At the exit of the first FPC, the divergence doesn't significant increases (δα < 1mrad) while the divergence has increase up to 3mrad after the second FPC. We can note an asymmetry in the first emittance diagram after the first FPC which results from the beam space charge expansion on one side of the beam.

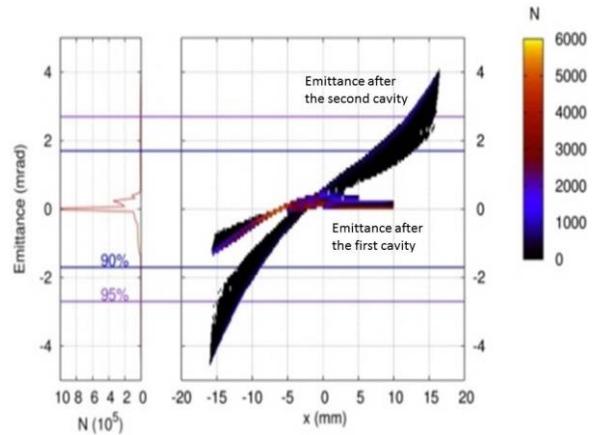

Figure 3.7: Beam emittance diagram at the neutralizer exit for the neutral beam. The left panel indicates the angular distribution of the beam particles and the horizontal bars accounts for deviation range corresponding to 90% and 95% of the total beam particles.

**Simulation with 60% photodetachment per cavity:** The total neutralization is even closer to the theoretical value of 84%; the simulation points out that increasing the laser power in the cavities (the photodetachment rate) reduces the beam emittance growth due the space charge effect with less than 0.8mrad after the first cavity and less than 2.5mrad after the second cavity. As a partial conclusion of this studied test-case supposing no residual gas in the neutralizer cell, the beam induced divergence in the photoneutralizer decreases with increasing the photon power in the FPC. The simulations of the secondary plasma formation which would cancel the beam space charge (the beam expansion in the photoneutralizer cell) with the residual gas (or a minor gas injection) in the photoneutralizer is part of further research combining OBI 3 [48] and ONAC.

### 3-3) Energy recovery system

The concept of the energy recovery is based on the conservative property of the Coulomb (electrostatic) force; the residual negative ions at the neutralizer exit which are at the high energy (1 or 2MeV) are decelerated down to a low energy (~100keV) and collected onto the recovery electrode to a potential close to the source potential, i.e., 100kV for a 1MeV beam (see Fig. 1.3), or -900kV for a 2MeV beam (see Fig. 1.4). In order to prevent the retro-acceleration of the secondary electrons emitted by the ion bombardment (on the recovery surface) towards the photoneutralizer at +1MV, the negative ions have to be collected within the recovery cell where a local vertical magnetic field ranging around 20mT deflects the negative ions towards the recovery surface and traps the secondary electrons within the chamber.

Fig. 3.11 shows a 3D simulation of the 2MeV beam recovery: the remaining negative ions (4A of $D^-$) at photoneutralizer exit are decelerated in two stages, from +1MV to 0V and to -900kV. They enter the recovery cell with an energy close to 100keV where they are transversally deflected by local vertical magnetic field (~20mT) and dumped onto the surface. The magnetic field is generated by two lateral coils similar to the Cybele ones (see Fig. 2.1-b) with an iron core which surrounds the recovery cell.

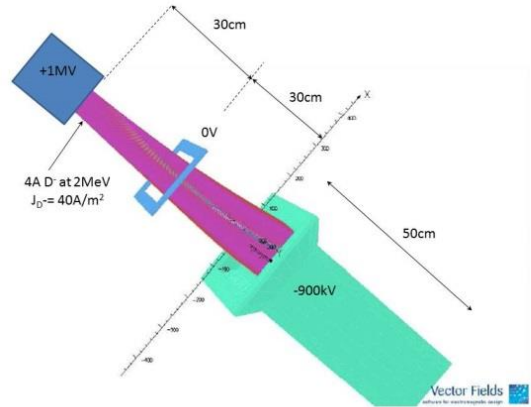

Figure 3.11: 3D simulation of the 2MeV beam recovery; the magnetic field is generated by lateral coils implemented onto the recovery cell (the coils are not represented on the figure), the field intensity is 23mT. The D⁻ beam is entirely collected and dumped within the recovery cell.

# 4- R&D IN PHOTONEUTRALIZATION

## 4.1 Photodetachment physics

The photoneutralisation-based neutral beam systems concept relies on the elementary process of photodetachment of a negative ion. A photon with an energy hν large enough can be absorbed by a D⁻ ion to make a detachment reaction occur:

$$D^- + h\nu \rightarrow D^0 + e^-.$$

Atomic properties, such as the electron binding energies and photoabsorption cross-sections depend on the atom mass only by a small isotope shift, which makes it easy to deduce how D⁻ photodetachment will work in a photodetachment-based high energy neutral (D⁰) beam injector from laboratory experiments carried out on H⁻

The photodetachment cross-section σ [9,49] is very low (only a few $10^{-21}$ m²) when compared to the few $10^{-12}$ m² possibly reached by the cross-section of an atom at the resonance of an electric dipole transition (such as a Na atom illuminated by yellow light at λ=589 nm), but photodetachment is a non-resonant process. This has, nevertheless, a positive counterpart: the wavelength is not critical, and the wavelength λ=1064 nm of a Nd:YAG or Nd:YVO$_4$ laser, sitting close to the cross-section maximum, appears very convenient. At this photon energy, 1.165 eV, the photodetachment cross-section has been known to be about 3.6(2) $10^{-21}$ m², according to most calculations and to the last measurement (see below). This is slightly less than the maximum cross-section of about 4 $10^{-21}$ m², to be found at wavelengths close to 850 nm, or 1.46 eV. This slight reduction of the cross-section should not however have drastic drawbacks, if the achieved light flux is enough to bring photodetachment to its saturated regime.

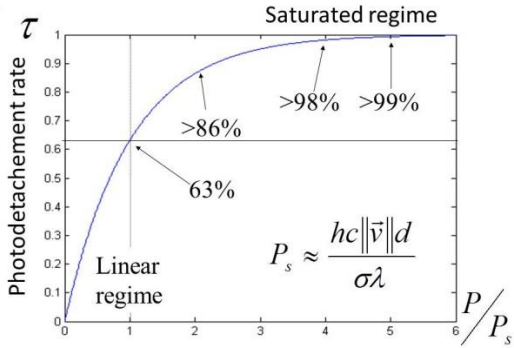 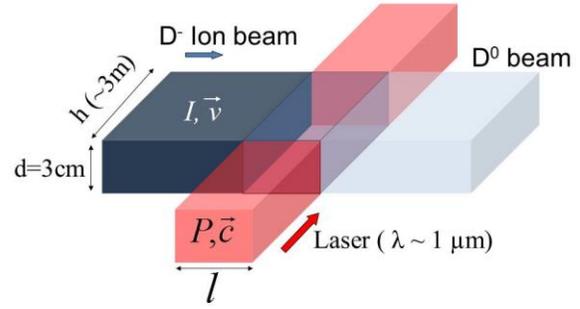

Figure 4.1: Photodetachment rate as a function of the relative photon power

Figure 4.2: Ion-photon interaction region

Another consequence of the non-resonant character of the photodetachment process is that photodetachment rate asymptotically increases with the photon power. The probability per unit of time of the photodetachment process being always the same, the population of ions illuminated by a laser beam decreases exponentially, without ever being reduced to zero (see Fig. 4.1). Yet after one "period" of the exponential decrease, it will already be reduced by a factor 1/e, which corresponds, as for the photodetachment process, to a 63 % efficiency. 95% efficiency is reached after only three "periods".

The value of the cross-section σ is a central parameter for dimensioning a photoneutralizer. The photon flux Φ necessary to have ions, of velocity v, reach one-period photodetachment decay rate of 63 % is just Φ=vd/σ, where d is the common transverse dimension of the overlapping ion and laser beams (see Fig. 4.2).

With σ≈3.6.10$^{-21}$ m$^2$ and a velocity v=9.8 10$^6$ m/s (1 MeV of kinetic energy, 2 MeV would increase it by √2), and d= 3cm, the required flux is just 8.2 10$^{25}$ ph s$^{-1}$, i.e., at the wavelength 1064 nm, of the order of 15 MW, which can be reached by four-or five times refolding of a 0.8 cm diameter 3 MW photon beam (see Fig. 1.1).

Before this project, the photodetachment cross-section of H$^-$ had never been measured with a laser. The last measurement of σ dated back to 1976 [49] and relied on the analysis of the radiation emitted by a hydrogen plasma. In 2013 one of our laboratories performed a laser photodetachment experiment using saturation of the photodetachment of an H$^-$ beam by a pulsed Nd:YAG laser to measure the cross-section without having to calibrate the ion current nor the detection efficiency [9]. The cross-section was found to be slightly larger than what most theoretical calculations had predicted, namely 4.5(6) 10$^{-21}$ m$^2$ instead of 3.6(1) 10$^{-21}$ m$^2$ [9]. A minority of calculations also predicts a higher value. A more recent and second laser measurement of the photodetachment finds the cross-section closer to the mean theoretical values [50]. The resulting uncertainty of the photodetachment efficiency in the SiPhoRE concept is not that large, anyway. If, for a given cross section the flux and interaction time lead to a 80 % photoneutralization efficiency, reduction of the cross section by 10% only reduces the efficiency to 76.5 %, i.e. by 4.4 %. Reduction appears even limited to 2.9 % if one has reached a 90 % efficiency. To be conservative, estimation of the necessary illumination must be made assuming the lower, $3.6 \; 10^{-21}$ m$^2$ value of the cross-section. Provided one has entered the saturation regime however, the ±10 % uncertainty that remains on the cross-section appears of smaller consequence than all other uncertainties concerning the implementation of a medium-finesse cavity on an ion beam, the performance of which could still vary within much larger limits.

**4.2  Towards a 3MW Photoneutralizer**

Although difficult, coupling a low power continuous-wave (CW) laser in a high finesse optical cavity, is not challenging in a breadboard based pure optics experiment in laboratory. So far, cavities with finesse as high as 2 million have been demonstrated [51].

Involving large Gaussian modes, long baseline cavity finesses are limited by scattering losses on the mirrors surfaces defect. Km scale cavities are currently operating in interferometric Gravitational Waves (GrW) detectors (Virgo, LIGO) [52,53], with an effective finesse around 2500 corresponding to a stored photon power in the 30kW range for 17 W of injected laser power [54]. A second generation GrW detectors (Advanced Virgo and Advanced LIGO projects) is expected to reach the MW range of stored power in the next few years [55,56].

GrW detectors cavities are faced with environmental seismic noise on which several resonances of the system structure and building are added, with frequencies lying beneath the kHz range. These mechanical noises can be highly reduced by the use of high performance vibration isolation systems called super-attenuators [57]. In addition, several servo loops are combined to tune and lock the input laser frequency on the cavity resonance (see for example [58]). These high bandwidth (several hundreds of kHz) feedback technics are robust regarding the mechanical vibrations in the laser beam direction and maintain a stable and steady state resonance [59] within the cavity.

In parallel to the cavity developments, it is worth to note that the specifications on the input CW laser are also severe: i) a well-defined Gaussian single mode has to match with the cavity mode; ii) a single frequency with a narrow bandwidth ($\delta\nu$ laser $\ll \delta\nu$ cavity = 150Hz); iii) a photon power higher than 500W.

None of the existing laser sources can guarantee such a high power for a single mode and single frequency beam. However, the achievements over the last decade of CW laser are significant; several hundreds of watts of single mode and single frequency laser beam have already been demonstrated in a fiber amplifier based system [60,61], making us confident on the fact that the kW range could be reached using coherent superposition technics on a few lasers beam [62].The appropriate solution would be to use a low power laser which is amplified in a high power optical fiber such as the Master Oscillator Power Amplifier (MOPA); it is a promising technology able to presently provide 100 W in CW without adding any overwhelming phase or amplitude noise [63].

At this point, it is worth to note that a 3 MW photoneutralizer for a Fusion reactor will fully take advantage of all these advanced technologies and experience gained on GrW detectors and CW lasers with the conditions that, first, the optical components of the photoneutralizer can be located sufficiently far away from the ion beam and the nuclear island of the reactor in a radiation shielded high vacuum vessel [10] (see Fig. 1.4) to avoid the deleterious effect of pollutants and radiations, and, second, the thermal effects on the mirrors which will play a central role with a 3MW photon beam can be overcome.

The 3MW cavity design consists in a four times refolded cavity with arms 25 m long (see Fig. 1.1) leading to a 100 m long cavity. All cavity mirrors are flat except the end one with a curvature radius of 1 km, which leads to a good optical stability less sensitive to misalignments and mechanical vibrations. Mirror alignments have to be under an active controlled via piezoelectric actuators in order to store the photon within the cavity. The intra-cavity photon beam width is 1cm with less than 10 % variation along the cavity. An in-depth description of the optical parameters of the cavity and of the active control of the mirrors being outside the scope of this paper, we only focus below on the main set of factors which appears the most important for the present feasibility study.

The intra-cavity photon ($P_{int}$) power is defined by $P_{int}= S.P_0$, where $P_0$ is the laser photon power and S the enhancement factor of the cavity.

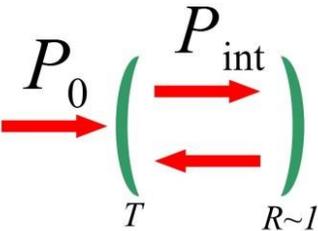

$$S = \frac{2F}{\pi\left(1 + \frac{\Lambda F}{2\pi}\right)^2}$$

$\Lambda$ represents the photon losses over one round trip in the cavity; it results from different sources:

- i) the scattering losses on the mirrors ~10 ppm per mirror for a beam width of 1 cm, i.e., 80ppm per round trip (see Fig. 4.5). This depends on polishing and coating technologies

- ii) Absorptions on the mirrors coatings ~0.3 ppm, i.e., 3ppm per round trip

- iii) Absorption by the ion beam ~3 ppm

leading to an enhancement factor S= 6300; as a consequence, 3 MW intra-cavity power could be reached with a finesse F=10000 and with an input laser power of $P_0$=500W.

### 4.3 ) Thermal effects on the mirrors:

With 3 MW of photon power on the mirrors, about 1 W is absorbed by the coatings (~0.3ppm per mirror), which gives rise to significant temperature gradients and thermo-elastic distortions. This is expected to distort mirrors surfaces in a non-spherical way leading to scattering losses and a drastic reduction of the cavity enhancement factor (S< 500).

Different solutions have been investigated like:

- Cool the back side of the mirror. This solution can only be applied to totally reflective mirrors and as such is not effective for the input mirror.

- Compensate the distortion by applying a mechanical stress.

- Compensate the mirror distortion by heating the back face of the mirror in order to achieve a compensating distortion

The thermal analysis described below considers a silica mirror of radius a=4cm and width h=5mm illuminated by a photon beam of radius *w*=1cm with an absorption rate of 0.3 ppm (~1W of thermal power absorbed).

The first step is to calculate the steady state temperature distribution within the mirror using COMSOL software. The second step is to compute the thermo-elastic displacement field. We concentrate on the z component (the direction of the light propagation) of the displacement field $u_z$ (r,θ) on the intra-cavity mirror surface, where (r,θ) are the polar coordinates of the mirror surface. The mirror is considered as a fused silica cylinder maintained on its edge. It receives a heat flux on its intra-cavity side corresponding to the absorption of the laser beam modeled by a Gaussian shape heat flux. At the wavelength corresponding to the temperatures considered here, silica is opaque and can be considered as a black body. The heat exchange with the external environment is made through radiation between the mirror's surface and the vacuum vessel with a uniform and constant temperature $T_0 = 300K$.

The obtained displacement field $u_z$ on the intracavity mirror side is then used to determine the residual optical behavior of the mirrors.

The Thermal Compensation System (TCS) proposed here involves heating the central part of the back side of the mirror with a heater at a temperature Tc and diameter b (see Fig. 4.7). The calculation has been performed for different values of the heater diameter b and temperature.

Without any correction system, the longitudinal temperature gradient induces a mirror expansion oriented towards the intra-cavity side (see Fig. 4.8 – Without TCS). With a central heater at a temperature Tc = 690 K and h=1cm (see Fig. 4.8 With TCS), the expansion takes place at the back side of the mirror (outside the cavity), while the distortion on the cavity side is nearly paraboloid.

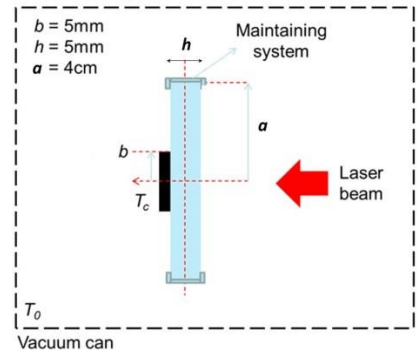

Figure 4.7: Thermal Compensation System with central heater implemented on the back side of the mirror (outside the cavity). The heater radius is 5mm, the mirror thickness is 5mm, and the intra-cavity photon beam power (laser beam) is 3MW.

An estimation of scattering losses is given by:

$$\Lambda = 4 \left(\frac{2\pi}{\lambda}\right)^2 \sigma^2 \qquad \text{where } \sigma \text{ is peak to valley distortion.}$$

The important peak to valley distortion observed without heater (see Fig. 4.9 –Without TCS) is considerably reduced (by a factor 20) with TSC leading to a much lower scattering loss (reduction by a factor 400).

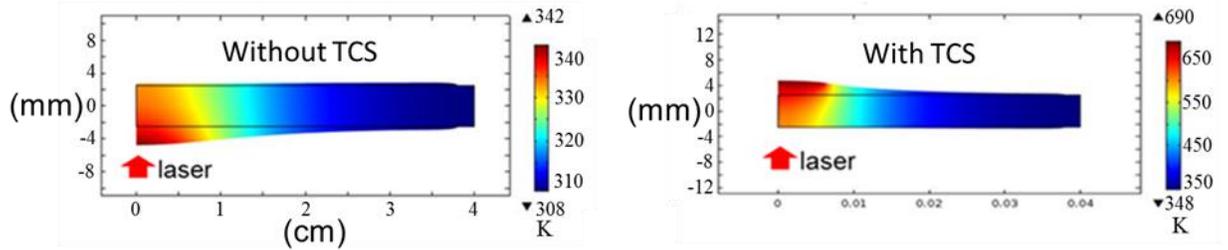

Figure 4.8: Mirror temperature field and distortion with/without compensation system.

With the central heater, the cavity enhancement factor is kept higher than 5000 corresponding to an input laser power of $P_0$= 600 W for 3MW of intra-cavity power.

Moreover, at the minimum of the scattering losses (Tc=690 K), the radius of curvature has changed from R = 1 km (initial value) up to L≃ 5.8 km. This central heating is only valid for the totally reflective mirrors since the TCS acts on the outer face of the cavity. Further investigations are to be conducted: (i) for the case of input mirrors for which the central part must be kept clear (see Fig. 4.5); an indirect heating via a $CO_2$ laser beam could be an interesting alternative. Moreover, we note that the scattering losses estimated above represent an upper losses limit within the cavity. Indeed, the scattered light from one mirror can be re-coupled to the cavity mode when reflected by another mirror.

Figure 4.9: Mirror residual distortion corresponding to the thermal distortion from which the spherical term is subtracted. This is responsible of the scattering losses.

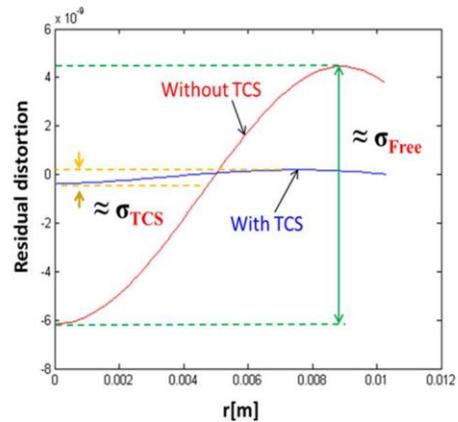

## 4.4: Photoneutralization experiment in laboratory

Intra-cavity photodetachment was used in some atomic or molecular physics experiments, either to produce neutral beams for collisional studies [64] or to enhance the photoelectron signal [65,66,67]. This was still far from saturation; the photodetachment efficiency in these experiments never exceeded a few %. In order to address the different issues encountered when mounting a higher-finesse optical cavity around a negative ion beam, the photodetachment microscope testbed at LAC has been modified to implement a Fabry-Perot optical-cavity finesse 3000 optical cavity (with 10kW intracavity power) able to achieved a nearly total detachment of a mm-size, 1 to 2 keV negative ion beam.

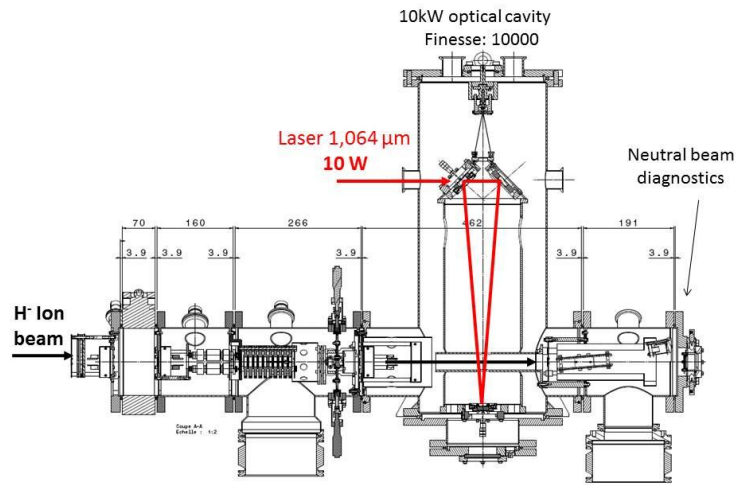

Figure 4.10: The photodetachment experiment at LAC equipped for intra-cavity photodetachment studies (Engineering drawing of the experiment).

It is a recycling triangular optical cavity composed of 3 mirrors (see Fig. 4.10) which is suspended within the tank to minimize the mechanical vibrations. To lock the resonance, two methods will be tested, i.e., the feedback on the laser wavelength which follows the cavity length fluctuations such as used interferometers, or implementing a cavity-length servo-locking procedure. At the time being, the optical cavity is under commissioning, it will be implemented in the vacuum tank in the near future.

## 5- HIGH VOLTAGE HOLDING IN VACUUM

As one of the goals of SIPHORE is to allow the DEMO neutral beam system to accelerate beams to 2 MeV, via -1 MV, 0 and +1MV potentials, voltage holding in vacuum is an issue. The current design of the ITER bushing that must provide the voltages to the 1 MV Mamug accelerator (via 5 accelerating grids) is of considerable size and expense and is also untested. Because the accelerator for SIPHORE aims to accelerate the beams over a single gap, the bushing can be simplified [14] because there is no need to provide five intermediate voltages through a single bushing.

This development aims to demonstrate a simplified compact bushing (see Fig.5.1) in order to hold the 1 MV [14]; an estimated 5 insulators are required that are separated by metal flanges. A total of 10 insulators are used because the bushing is a closed gas-tight system and 1 MV insulation must be provided between the SF6 and the vacuum side of the beamline vessel. The gas-tight bushing allows its interior to be filled with low-pressure gas (~0.01 to 0.05 Pa) that serves to suppress dark current [14,69]. This dark current is an electron current, it results from an electron field emission process [68] which occurs on the cathode metal surfaces in a vacuum system. Before to design and develop a Bushing at scale one, the ongoing experiments target to find the best high voltage holding conditioning protocol under vacuum between two cylindrical electrodes representing one 200 kV Bushing stage, the main goal being to minimize the dark current flowing between these electrodes under a high electric field (~50 kV/cm or more).

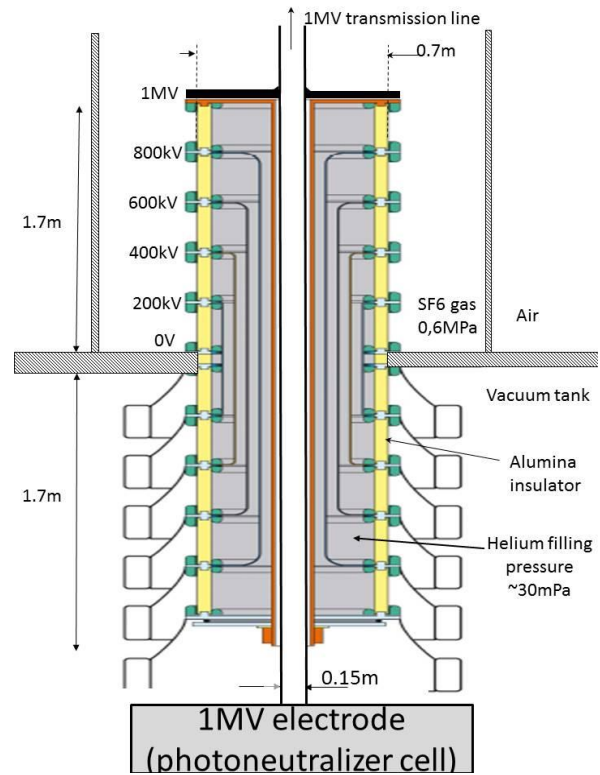

Figure 5.1: 1MV compact Bushing concept.

It can be seen that when gas is added to the vacuum vessel, the emitted current is significantly reduced. With N2 and Ar efficient dark current suppression (more than 90%) occurs with a time constant of the order of 1 minute.

While designing and procuring a single stage test bushing, experiments dedicated to test the resilience of several electrode materials (Cu, Ti, Mo, stainless steel) against high stored-energy breakdowns have been performed. Little difference was found, which means that the cheapest material (stainless steel) is a good choice [70]. Implementation of the test bushing on the testbed is now finished and experiments have started.

Design, procurement and installation of a 2-stage 400 kV bushing are expected to finish in 2016. This bushing should incorporate best practices in terms of surface preparation and gas found during the preceding work. If testing of this bushing is successful, one can consider proceeding with a design for a 1 MV bushing. In Fig. 5.2 the effect of different gases on the electron emission (dark current) generated between electrodes [69] is shown.

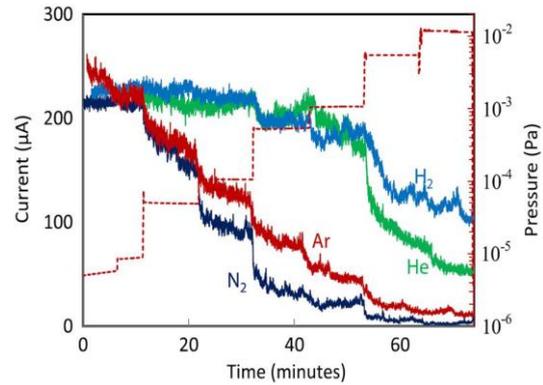

Figure 5.2: Field emission current decrease for four different gases with increasing gas pressure: 2 cm gap distance between electrodes, 30 kV applied voltage. Figure reproduced from reference [69].

## 6- SUMMARY AND CONCLUSIONS

The next generation of fusion reactors (DEMO Tokamak) requires a high level of additional plasma heating power ranging between 100 and 200 MW. The overall efficiency of the heating systems has to be high enough ($\eta > 60\%$) to reduce the recirculating electrical power within the plant to an acceptable level. This paper presents in the first part (section 1) the concept and potential performances (high efficiency and high neutral power) of a new neutral beam system which is based on the photodetachment of the energetic negative ions ($D^-$). The two centrepieces of the system are the "Photoneutralization" and the "beam recovery" which lead to very low beam losses and a consequent high injector efficiency (>60%) with more than twice the neutral power produced in respect of conventional NB systems that are based on a gas neutralizer. The second part of the paper (sections 2 to 5) presented the concepts and the accompanying R&D around the ion source, the accelerator, the photoneutralizer and the high voltage bushing; these are all challenging issues.

At present we can put forward some partial answers to the Siphore concept feasibility:

-i) Section 2: the first results of the ion source (experiments and plasma modelling) show that the magnetic confinement enables the achievement of a long and thin uniform plasma with a radial plasma cooling which should favor the production of negative ions in the extraction region. Doped diamond at an operating temperature close to 500°C produces a significant negative ion yield by respect to other tested materials (HOPG, CFC, Mo, Cu). Further experiments will be performed testing different kinds of materials and diamond monocrystals with varying boron (and nitrogen) dopant rates. The objective being to understand surface-production mechanisms and to qualify the materials (long term stability, sputtering, contamination, etc..) under plasma exposure.

-ii) Section 3: the preliminary 3D negative ion beam simulations of the accelerator, photoneutralizer and recovery systems have been performed regarding the production of an intense blade shaped negative ion beam at 1 or 2 MeV. Further beam optics simulations and modelling are required to optimize the beam divergence and aberrations for a high neutral transmission in the duct, the target being an aberration-less ion&beam with a divergence lower than 5mrad.

-iii) Section 4: the keystone and main challenge of the injector is the achievement of a high power Fabry-Perot cavity which has to provide a 3MW photon power in continuous wave.

The thermal studies of the mirrors which will play an central role in such a system highlight the deleterious effect of the photon power absorbed by the mirror coatings ( ≈0.3ppm per mirror equivalent to 1watt) which gives rise to significant scattering losses (>1000ppm per roundtrip). Conversely, a mirror correction based on a central heater set up on the mirror back face makes it possible to reduce the scattering losses to less than 3 ppm per roundtrip. Further in-depth mirror thermal studies will be performed, the goal being to find a technically feasible solution combining low scattering losses with high cavity stability. In parallel, a first small-scale experiment of negative ion beam photoneutralization in a cavity (10 kW of intracavity photon power) is under commissioning to give partial answers on the photoneutralization feasibility, such as the effect of the beam in a cavity, the mirror pollution and their lifetime in the beam environment, the effect of the mechanical vibrations, the servo loop for the lock of the cavity resonance.

-iv) Section 5: R&D around a simplified compact bushing able to hold the 1 MV with low dark current on a Siphore system is ongoing; experiments on a small-scale testbed show that when gas is added to the vacuum vessel, the emitted current is significantly reduced. With $N_2$ and Ar efficient dark current suppression (more than 90%) occurs with a time constant of the order of 1 minute. In parallel, as a premise to the bushing development, full-scale experiments on dark current and voltage holding are performed on a single-stage 200 kV bushing under different electrode surface physical parameters to find the best high voltage conditioning protocol.

To varying degrees, each of these different research fields will allow to assess in detail the overall feasibility of the system before considering the RAMI (Reliability, Availability, Maintainability, Inspectability) issues and its implementation in the environment of the fusion reactor where other open challenging questions, such as "Remote handling", will have to be addressed.

## ACKNOWLEDGEMENTS

This work benefited from financial support by the French National Research Agency (Agence Nationale de la Recherche; ANR) in the framework of the project ANR-13-BS04-0016-04 (Siphore), ANR-13-BS09-0017-03 (H-Index), ANR- 12-BS09-013 (HVIV), ANR-11-JS09-008 (METRIS) and by the PACA county in the framework of the project "dossier 2011_11042" (PACA Siphore) and "dossier 2012_10357" (PACA-Ging). This work has also been carried out within the framework of the EUROfusion Consortium and has received funding from the European Union's Horizon 2020 research and innovation programme under grant agreement number 633053. The views and opinions expressed herein do not necessarily reflect those of the European Commission. This work was also supported by French Government through a fellowship granted by the French Embassy in Egypt (Institut Français d'Egypte).